# Beyond source attribution: Address authorship and comparability decisions in an LLM-built scholarly knowledge graph

(Companion to https://paragraph-lab.org/casebook)

Dr Ksenia Kozha
*ksenia.kozha@aucy.ac.cy*
PaRAGraph Research Project[1]

ABSTRACT:

A knowledge graph needs a stable address for its evidence even where the literature offers competing classifications and agrees on none. Yet an address chosen for retrieval readily passes for the only classification available. This happens whenever records gathered under one name concern units that differ in boundaries, rank, or the criteria by which they are described or counted. In Sinitic dialectology the same localities recur in schemes whose units differ in membership and rank, and a unit may change its affiliation when the evidence for its diagnosis changes. Two tone counts, each quoted faithfully, may still count different things. The taxonomy of an entire field thus comes to rest on the unstable frame of a literature that contradicts itself. Neither verifying a quotation nor searching for conflicting values detects these differences. I distinguish the working address that gathers candidate evidence, the category that the source defines and the decision that admits a passage to a particular comparison, and argue that such decisions, together with the dependence among records, belong to the provenance of a comparison. On four re-runnable cases, the accompanying *PaRAGraph Casebook* shows where such differences arise among claims extracted by a language model and verified before entering the graph. The platform now stores the corresponding decisions as objects for one dimension of comparability, the address of a record: the placement of a locality in each source's scheme, the decision that a name in a given publication denotes a unit rather than a place, and the origin of every address. The working taxonomy can then be revised as new sources yield new claims, while the path from each comparison to its evidence stays fixed.

*Keywords*: knowledge graphs; source attribution; classification; claim extraction; comparative concepts; Sinitic dialectology; Yue

[1] *PaRAGraph Research Platform* for Sinitic dialectology offers validated methodology for source-grounded knowledge graph construction from academic corpora, with measurable precision and recall baselines for LLM-assisted claim extraction.*:* https://paragraph-lab.org/

# 1. Introduction

The second edition of the *Language Atlas of China* divides Yue dialect group into seven piàn 片 'subgroups', among them Gaoyang 高陽 and Wuhua 吳化 (Zhang 2012: 127). Wuchuan 吳川 and Huazhou 化州 figure in the membership of both, the overlap being resolved only below the level of the county (Gaoyang claims part of Wuchuan and the eastern part of Huazhou, whereas Wuhua restricts each unit to named townships). An earlier division, which had assigned Wuchuan, Huazhou and Zhanjiang to two piàn at once, drew the objection from Anne Yue-Hashimoto 余靄芹 that the double assignment was nowhere explained (Yue-Hashimoto 1991a: 166–167).

In print, an author can leave such a classification unresolved, acknowledge the dispute in the main text, or discuss it in a footnote. A knowledge graph has no footnotes. To retrieve evidence about the area it must assign it an address, and an address indispensable for retrieval readily becomes an unacknowledged premise of comparison: two publications filed under the same subgroup appear to describe a single unit even when they define it with different members and at different ranks.

Disagreement in a claim graph is most simply detected as a conflict of values. Classificatory disputes rarely take that form, for they concern which units exist, where their boundaries run and at what depth of the hierarchy they sit. The question addressed here is what must remain open to inspection when a researcher compares records retrieved through such a taxonomy, or replaces that taxonomy with another. The answer I propose distinguishes three things: the working address under which the system assembles candidate evidence, the source category that a publication actually defines, and the applicability decision by which the researcher admits a passage to a particular comparison. It is this last decision that makes two records comparable or keeps them apart, and it is the one that current representations leave unrecorded. Source-qualified concepts and attributed mappings between them are established techniques (Section 2). The question posed by extracted claims is narrower and comes earlier: whether a given record, reached under a given working address, bears on this comparison at all, and on what grounds.

The accompanying *PaRAGraph Casebook* publishes four versioned case bundles in which the problem can be inspected record by record. In C1 and C3, a search for opposite polarity finds no conflict although the sources disagree about the very units to which the records are addressed. In C2, a record lacking the relevant locality tag belongs in the comparison only because of context outside that record. In C4, a quotation supports a number whose unit, or whose dependence on a repeated table, makes it unsuitable for the intended comparison, and the record itself says so in free text. On these published records I show that the warrant for admitting a claim to a comparison cannot be recovered from the claim, its quotation or its value, and must therefore be stored as an object in its own right. The *Casebook* keeps such warrants as curatorial annotations beside records exported in September 2026. The production graph now stores them as objects for one dimension of comparability, the address of a record (Section 6). Extending the same mechanism to units of count, the scope of a passage and the dependence among records is the next stage of the work.

## 2. Related work

The representation of a concept as defined by a particular source is an established problem outside linguistics. In biological taxonomy the Taxonomic Concept Transfer Schema identifies taxon concepts by the publications in which they are circumscribed (Taxonomic Names and Concepts Interest Group 2005). The TCS 2 list of terms, released for public review in 2025, adds standalone mappings between concepts with their own source attribution (TDWG Taxon Concept Schema (TCS) 2 Task Group 2025), and Euler/X reasons over expert alignments across taxonomies (Chen et al. 2014). The present argument does not depend on treating source-indexed concepts or their mappings as new. Its locus is an earlier decision in an extraction workflow: whether a particular claim and its passage bear on a comparison reached through a working graph address, before any relation between the source schemes is asserted.

Within linguistics the nearest working solution is arguably Glottolog. Each release publishes a single classification, which its editors call their 'best guess'; a language whose closest relatives cannot be determined is 'left unclassified within the finest-level (sub)family that can be discerned' (Hammarström et al. 2026). Reclassification is traced through the assignment, retirement and versioning of Glottocodes (Forkel & Hammarström 2022: 921–922). Uncertainty is thereby described at the meta-level, while each languoid retains a single parent within any given release. PaRAGraph proceeds otherwise, storing competing classifications as separate attributed claims, none of which is promoted to parenthood over the others. From network approaches to Chinese dialect history (List 2015: 29–30, following Baxter 2006) I adopt not a method of reconstruction but a requirement of storage: every classificatory claim must retain its criterion, its material and its source.

The closest model for the atomic unit adopted here is claim-centred publishing. Nanopublications package an assertion with its provenance (Groth et al. 2010), and micropublications extend the model to evidence, arguments, annotations and challenges (Clark et al. 2014). PROV-DM supplies the general vocabulary of entities, activities and responsible agents in which such records are described (Moreau & Missier 2013). OntoDup goes furthest in this direction for scholarly knowledge graphs, recording the decisions by which duplicate bibliographic records are merged as reified assertions that carry evidence, provenance and review status (Galán-Mena et al. 2026). The decision examined here is of another kind: it concerns not the identity of two records but the bearing of one record on one comparison, and the ground on which that bearing is accepted. ProVe tests whether a knowledge-graph triple is supported by the text given as its provenance (Amaral et al. 2024), and work on attribution asks the same question of generated statements (Rashkin et al. 2023). C4 supplies a case in which the quoted text supports the extracted number but the record still cannot enter the proposed comparison without a further judgement about its unit and context. Attribution, in other words, is satisfied, while comparability is not.

On the linguistic side the problem is that of the relation between descriptive categories and comparative concepts. Categories established for the description of one language are not thereby identical across languages (Haspelmath 2010). The same caution holds within a single dialectological literature, where piàn, qū 区 'district' and xiǎopiàn 小片 'subdivision', or diàolèi 调类 'tone category' and diàozhí 调值 'tone value', derive their force from particular analytical traditions. Cross-linguistic databases define their parameters before any value is coded (Forkel et al. 2018). An extraction pipeline inverts this order, receiving values from heterogeneous descriptions first and leaving the parameter to be reconstructed from them.

## 3. *PaRAGraph Casebook*

*PaRAGraph* is a research environment built over a library of scholarly publications, with Sinitic dialectology as its first domain of validation. Documents are divided into text chunks mapped to page ranges, and each chunk into numbered sentence segments. A large language model reads one chunk at a time and proposes claims. Each claim names a feature drawn from a controlled vocabulary, a working target, a value where one applies, a polarity (attested, not attested or uncertain) and the conditions stated by the source. The model does not write its evidence. It points to a range of segments, from which the system cuts the quotation and its character offsets, so that the quoted text is always the text of the source. Claims are stored first in a relational database, where a second model and a set of deterministic checks score each one against its passage. Claims above a threshold are promoted automatically to a Neo4j graph whose nodes represent groups, subgroups, clusters, features, sources, claims and evidence; borderline claims wait for a curator. Every evidence node keeps the identifier of its chunk, so that each record can be traced to its passage, and nothing enters the graph from outside the library. The graph can be queried directly in Cypher, as in the Casebook, or through a chat agent whose graph component returns claims grouped by feature and target together with the passages used for citation.

Several design choices matter for the cases. The working target is invariably a group, subgroup or cluster of a managed taxonomy. For Yue the subgroups follow the seven piàn of the 2012 Atlas, adopted as a convenience of retrieval and not as an endorsement of that classification. A locality below the level of the subgroup is represented not as a node but as a normalised tag on the claim, together with the rule that mapped it, so that a single county may fall under different targets in different claims. It is in this way that 廉州 Lianzhou in Guangxi (YUE_QINLIAN) is kept distinct from 連州 Lianzhou in northern Guangdong, whose claims are filed under the Tuhua subgroup TUHUA_YUEBEI, although the two names are identical in romanisation. Polarity records what the source asserts, not the confidence of the model, which is stored separately. A reported absence therefore becomes a claim and not a gap in the data. Coordinates, where a locality has been geocoded, are held in a separate locality registry together with the placements of that locality in different classifications, each marked by its origin. The position of a county on the map is thus kept apart from the subgroup to which a source assigns it.

*The Casebook* (Kozha 2026) is an offline viewer with one tab for each case, running on bundled JSON without live API calls. Each case has a hash-checked bundle of exported records and explicitly marked curatorial annotations. The means of examining those records differ across the cases. For C1–C3, native operations (Cypher queries over graph fields) were re-run against the production graph on 10 September (C1), 9 September (C2) and 11 September 2026 (C3). C4 uses a frozen export dated 16 September 2026, checked against the live graph on 20 September; its extraction SQL and reviewed subset are retained in the bundle, alongside proposed operations and editorial observations. All figures in Section 5 refer to these dated artefacts; the production architecture described in Section 6 is given as of 3 October 2026.

## 4. Three levels of representation

*The Casebook* analysis distinguishes three levels of representation. A working address is the label under which the system retrieves candidate evidence. A source category is the unit or analytical category that a publication itself defines. An applicability decision records the grounds on which the researcher treats a passage as relevant to a stated comparison.

With respect to the classified object, YUE_GAOYANG is a working address. The source may define a subgroup (*piàn*), a district (*qū*) or a subdivision (*xiǎopiàn*), each with a membership of its own, and it falls to the researcher to decide whether such a unit corresponds to one in another scheme. With respect to the feature, TONE_INVENTORY_COUNT is likewise a working address. The source may count tone categories (*diàolèi*), tone values (*diàozhí*) or merely the columns of a table, and it is again the researcher who decides which values admit comparison.

In *the Casebook* each comparison is documented as a route from a research question through selected statements and their surrounding passages to the curator's inclusion rationale and interpretation. The route also records dependence where it can be established, since records, evidential passages and independent source analyses are distinct quantities. Two records may repeat a single table header, and two publications may transmit one earlier proposal.

| Case | Dimension | Observed difficulty | Decision |
|---|---|---|---|
| C1 One territory, four partitions | boundaries | The same place names occur in differently defined units; lists overlap within a single source | Keep membership, local qualifiers and attribution; assess equivalence separately |
| C2 Same variety, new evidence | evidential basis | A classification changed because the evidence changed; one record applies only through context | Keep the basis of each statement, the contextual decision and shared support |
| C3 Same names, different depths | rank | Equal names occupy different levels; the only cross-scheme relation is approximate prose | Keep rank, parentage, edition and the stated scope of correspondence |
| C4 Same field, different units | unit of count | Values mix categories, contrastive units, partial tables and repeated headers | Record the unit and evidence type; count analyses, not records |

Table 1. *The four Casebook cases and the requirement each yields.*

## 5. Four cases

### 5.1. C1. One territory, four partitions

C1 concerns south-western Guangdong around Wuchuan, Huazhou, Zhanjiang 湛江, Yangjiang 陽江 and Yangchun 陽春, an area organised differently in four published positions. The Atlas recognises two piàn here: Wuhua includes named townships within Potou district of Zhanjiang and within Wuchuan and Huazhou, whereas Gaoyang lists eleven county-level units, including the eastern part of Huazhou, part of Wuchuan and most of Lianjiang (Zhang 2012: 127). Wuchuan and Huazhou thus appear under both Atlas units with local qualifications, and the lists do not state a disjoint partition. An Outline of Yue Dialects in Guangdong unites Wuchuan, Huazhou, Yangjiang and Yangchun in a single Gao-Lei 高雷 piàn (Zhan 2002: 1, 5). Yue-Hashimoto works with a different system of ranks, comprising two qū, five piàn and four xiǎopiàn (1991a: 176). In her scheme Huazhou and Wuchuan belong to the Interior subdivision of the Northern Delta piàn (1991a: 178), Yangjiang and Yangchun constitute a Liangyang 两陽 piàn, and no unit corresponds to either Gaoyang or Wuhua. Zhang & Zhuang (2008: 421), finally, propose a Liangyang piàn for the Moyang basin and a Gaohua 高化 piàn for the Yue spoken south and west of Xinyi and Gaozhou.

The working addresses YUE_GAOYANG and YUE_WUHUA bring the relevant records together but cannot establish that the named units are identical; rank belongs to a source's scheme, not to the working address. Fourteen statements from five sources rest on thirteen distinct evidence nodes, since the assignments of Wuchuan and of Huazhou share one node. These are textual supports, not thirteen independent analyses: Zhang and Zhuang, for one, discuss the Atlas and the Outline explicitly. A query for opposite polarity in this selection returns no rows. The dispute about membership and rank stays invisible to it, because nothing in the ontology encodes that dispute as a polarity conflict.

Reading the Chinese evidence against the stored values brings out a further class of error, which the published bundle allows any reader to repeat. The Atlas list for Gaoyang names eleven county-level units, seven of them qualified by markers such as 大部分 'most of', 部分 'part of' or 东部 'the eastern part'. The English value written by the model names nine, omits the districts of Maonan and Jiangcheng, strips the listed units of their qualifiers and renders 安铺等大部分地区 'Anpu and most other areas' as 'Anpu and nearby areas'. Records in the same bundle whose value kept the Chinese list are correct. Paraphrase of an enumeration loses members and markers of partiality, and for claims of distribution each such loss may create or erase an overlap between units. The native value is published as exported, next to its quotation; the discrepancy is established by reading the source, not by any operation over the record.

## 5.2. C2. Same variety, new evidence

C2 concerns Lianzhou 廉州 in Hepu, Guangxi. Yue-Hashimoto (1991b: 299) reports aspirated reflexes of the earlier voiced stops and affricates irrespective of tone, a pattern which, taken as the sole criterion, would place Lianzhou with Hakka or Gan. Later in the chapter she records its earlier identification as Hakka and argues for its Yue affiliation on tonal and lexical grounds (p. 311). Footnote 21 attributes the earlier view, advanced in her own work of 1970, in part to the material then available, character-based field notes designed to establish phonological correspondences, and reports that subsequent fieldwork on colloquial speech confirmed the Yue status of Lianzhou (p. 319). The classification changed because its evidential basis changed, not because the variety did. A graph retaining only the final affiliation would register no more than a change of label, whereas the Casebook preserves the criterion, the material and the history of revision.

The case also exposes a weakness in the graph's own vocabulary. The feature CONS_VOICED_OBSTRUENTS_REPORTED receives both statements about voiced consonants in a modern inventory and statements about the reflexes of the Middle Chinese voiced initials. The source keeps the two levels apart; the vocabulary does not. Across the wider Qinlian target a flag in the stored value happens to separate them in all fifteen records, but nothing in the schema requires it, and the wording that distinguishes a historical reflex from a synchronic inventory survives only in free text.

The worked applicability decision concerns S4, a subgroup-level record that bears no locality tag but whose source context names Lianzhou in the preceding table. As curator I admit it on that ground, and the view grows from eight records to nine. The added record widens the admissible evidence without constituting a new observation. Dependence is concealed here as well. Nine statements from two sources (Yue-Hashimoto 1991b; Chen & Chen 2005) rest on seven independent textual supports, and shared basis can be detected only by comparing evidence spans, since a test confined to the same chunk would wrongly pair two adjacent passages. The single contested cell, concerning voiced obstruents, reflects an inconsistency of coding rather than a dispute between scholars: two records of the same source carry the same negative value under opposite polarity.

## 5.3. C3. Same names, different depths

C3 compares two schemes of unequal depth. The Atlas divides Yue once, into seven coordinate piàn such as Siyi 四邑, Gaoyang and Guangfu 廣府, and posits no unit between Yue and the piàn (Zhang 2012: 127). Yue-Hashimoto's Table 5, by contrast, has separate columns for qū, piàn and xiǎopiàn (1991a: 176). Her Siyi–Liangyang district contains a Siyi and a Liangyang piàn, and her Delta district contains the Northern and Southern Delta piàn together with a Guangfu piàn. The Appendix to Development of the Stop Endings in the Yue Dialects presents a related scheme under headings such as 四邑兩陽區四邑片 and 三角洲區廣府片 (Yue-Hashimoto 2002: 241–242). Its memberships and the name of one subdivision differ from Table 5, so the Casebook keeps the two as separate witnesses. Siyi is a piàn within the Siyi–Liangyang district, while Guangfu belongs to the Delta district. The identically named Atlas units are coordinate divisions directly under Yue: a shared name establishes neither rank across schemes nor parentage within one.

The only sentence in the bundle that relates the two schemes is supplied by Kwok, T'sou & Chin (2015), for whom Yue-Hashimoto's Siyi–Liangyang 'roughly corresponds to the Sìyì sub-group and parts of the Gāoyáng sub-group' of the Atlas. The correspondence is explicitly approximate and refers to the first edition of the Atlas (Wurm et al. 1987) rather than the second. It is, moreover, carried by no graph record, because the adjacent extracted claim ends one sentence earlier, and the Casebook retains it as a curatorial annotation. Liangyang cannot therefore be read as another name for Gaoyang, still less so since the Appendix assigns to it Hecun in Xinhui, which lies within the Siyi territory of the Atlas.

A scheme of this kind exceeds what the working taxonomy can hold. Six of the seven records drawn from Yue-Hashimoto 1991a are attached to the group address YUE, since a source that describes how a language is divided offers no subgroup to which its statements could be attached. The graph has no address for Liangyang, which is reachable only through its Siyi piàn, and Kwok, T'sou and Chin's report of the 1987 Atlas is filed under the feature CLASSIFICATION_ATLAS2012, an identifier that dates the wrong edition. Units of this kind therefore have to be recorded apart from the working taxonomy, as decisions attributed to the publication in which they occur; this is what the production graph now does (Section 6).

## 5.4. C4. Same field, different units

C4 asks whether the argument extends from classification to a feature, and its question is elementary: what does a number stored under TONE_INVENTORY_COUNT count? The public C4 bundle contains a frozen export of 504 Yue records from 29 documents, dated 16 September 2026. Its three-variety comparison view holds 65 records, grouped into 17 locality–document–value pairs and tagged to Xinyi 信宜 (29), Taishan 台山 (27) and Guangzhou or standard Cantonese (9).

The three varieties expose distinct counting problems: fragmented tone tables for Xinyi, recurring dictionary tables for Taishan and differing analyses of standard Cantonese. The Casebook retains editorial review labels: A marks a number with the correct basis of count, B a number that agrees with the author's but rests on an unsuitable basis, and C an extraction or chunking artefact. These labels describe the selected records, not a measured accuracy rate. Two questions remain open in the published Casebook: the adequacy of the quotation behind one standard-Cantonese record, and the attribution of two Xinyi records.

Numbers that appear contradictory may be compatible. Ye Guoquan 叶国泉 distinguishes nine tone categories (diàolèi) from eight tone values (diàozhí) for Xinyi (Ye 1987, in Luo 1987), and Huang Jianyun 黄剑云 five categories from eight values for Taishan (Huang 2017). For standard Cantonese, Bauer & Benedict count ten tone categories (1997: 121–122), but seven contrastive tones for speakers from Guangzhou and six for most speakers in Hong Kong and Macau (1997: 248–249). Conversely, identical numbers need not reflect identical analyses: Tang Zhidong 唐志东 and Wang Jianshe 王建设 both count eight categories for Xinyi but assign the tone with value 13 to different categories (Wang 2003; Tang 1984 as reported by Wang).

Two failures bear directly on the design of such systems. The first arises from segmentation. In the Xinyi volume the tone tables were divided so that six non-entering columns fall into one chunk and three entering columns into another. The model extracts six in 15 records and three in two, and a quotation-level check confirms each fragment. Six thus becomes the most frequent Xinyi value in the comparison view, although no source proposes six as its full tone inventory. The second failure arises from multiplying records. In the Taishan dictionary twelve records with the value five come from recurring tone tables in separate entries and chunks, and fourteen further records count indices within entries. Because each record sits in its own chunk, a query for shared evidence finds nothing to merge: the repetition belongs to the structure of the source, not to the storage of the graph.

What makes these failures instructive is that the model was not ignorant of them. In nine of the fifteen Xinyi records valued six, the record's own conditions state that the count covers only the non-entering tones ('six 舒声 tone categories … entering tones additional'; '不含入声'). In eleven of the twelve Taishan records valued five, the conditions restrict the count to rhymes without entering tones. Every one of these records passed verification and was promoted to the graph without human review. The scope of the number was written down, in a free-text field that no operation reads, while the scalar field carried the number on into comparison. The schema compounds the difficulty from the other side: a field constrained to one integer cannot express 'five categories and eight values', although that is precisely what the source states.

### 5.5. What the cases establish together

Difference at the level of representation and difference between scholars do not coincide. Substantial disagreement over membership and rank produces no record of opposite polarity in C1 and C3, whereas the only such pair in C2 arises from inconsistent coding; in C4 a single account yields different numbers, while different analyses yield the same one. Within these selections, conflicting values are neither a necessary nor a sufficient indication of disagreement between scholars.

Evidential dependence is concealed in the same manner, whether through an evidence node shared by two records in C1, a footnote shared in C2 or a header repeated across entries in C4. A count of claims may therefore overstate support even when every record can be traced to genuine text, and the proper unit of support is the source analysis together with its evidential basis.

Certain errors become visible only when a record is read against its source, among them a paraphrased list of localities and a feature identifier that dates the wrong edition. No operation over native fields can detect them, and *the Casebook* leaves the native records as exported so that the discrepancy remains open to anyone who reads the quotation. Two feature definitions also need revision: ‘voiced obstruents’ conflates synchronic presence with historical reflexes, and the tone-count field requires one integer where the source distinguishes two kinds of count.

The most consequential observation, however, concerns what was available rather than what was missing. In C4 most of the records concerned state, in the model’s own words, that the Xinyi six and the Taishan five exclude the entering tones; in C2 the wording that separates a reflex from an inventory is present in the quotations. The knowledge needed to decide comparability was in many cases already produced and was lost only because it had nowhere to go. This is the strongest argument the Casebook offers for treating comparability as data.

## 6. Comparability decisions in the graph

The production graph answers C1–C3 with an address-authorship layer, introduced between 30 September and 2 October 2026. Its first component records the placement of each locality in a classification as a row of its own, with an origin (the Atlas, a curator or an automatic agent), a locator in the source and a status: single, split or disputed. On 3 October 2026 the registry held 2,776 such rows. Wuchuan and Huazhou, the localities of C1, each carry two Atlas rows, under Gaoyang and under Wuhua, both marked split; Wuchuan has a further row under Leizhou Min, and Huazhou one under the Hakka of western Guangdong. When a locality is split or disputed, the system does not choose a subgroup on the source's behalf. Unless the passage itself names the unit, the address of the claim stays at the level of the group.

The second component is the curator's decision, stored as an object with its author, date, scope and record of application, so that it can be audited and rolled back. The position and level tables of *the Casebook* go one step further: derived from an offline test of a layer describing the source schemes themselves, they record each unit with the publication and edition from which it derives, its name and rank as written, its parent and its members. The third component marks every claim promoted under the current finalisation with the origin of its address (named by the source’s author in the passage, proposed by the model, derived from the project vocabulary or assigned by a curator) and with the level at which the name was found.

For numerical features the corresponding instrument is a tone passport, which records what was counted, which analytical decisions bear on the count (the treatment of entering tones, sandhi, the population described) and how the value was obtained. *The Casebook* includes 21 passport rows for the three varieties, of which 17 are linked to native claim IDs and cover 59 comparison-view records; four remain unlinked rather than being matched by conjecture. Ten linked rows are recovered through locality-tagged records, while seven require a source-and-value route through the subgroup address. The passports were prepared by hand. C4 suggests, however, that much of a passport could be proposed by the extraction itself, since the model already writes the unit and scope of a number in free text; the curator would then confirm a structure instead of composing it.

These components give the decision about an address the form that this article claims for comparability decisions in general: an object that links a record to a scheme, states its ground, names its agent and date, and can be applied, audited, superseded or rolled back without altering the claim. The same form extends to the other grounds the Casebook documents. For C4 such a decision would link a record to a comparison and state its unit of count; for C2 it would record the contextual locality that admits S4; others would record the scope of a quoted passage or the dependence of one record on another. In the terms of PROV-DM (Moreau & Missier 2013) each is an activity, attributed to an agent, that uses a claim and generates the membership of a comparison. What PROV leaves open, and what the Yue cases supply, is the vocabulary of grounds.

With decisions stored in this way, revision becomes an operation that can be tested while the evidence is held constant. A researcher replaces the adopted hierarchy, that of the Atlas for instance, with another, such as Yue-Hashimoto's, and examines which selections and conclusions change. The test succeeds if both interpretations are preserved, every affected comparison is identified and each result remains linked to its passages and decisions of inclusion; an approximate correspondence must remain approximate. Evaluation should record whether reviewers can recover the passages needed to verify a result, how often they agree on applicability and how long they take to reconstruct the evidential basis of a result. These are targets of evaluation, not reports of measured performance.

The same method could test the features used as classificatory diagnostics. Once observations from different publications have been made comparable, one could examine which features co-occur, which exclude one another and whether their distributions follow accepted subgroup boundaries or cut across them. Such relations would form a layer against which classifications are tested, so that a classification ceases to be a frame fixed in advance and becomes a hypothesis about the structure of the dialect space. Nothing in this is specific to Chinese. A ceramic phase in archaeology, an offence category in legal history or an occupational code in historical censuses raises the same questions of address, source category and applicability.

## 7. Limitations

The four cases are deliberately selected: they expose the mechanisms by which comparability fails, not the rate at which it fails, and error rates across the corpus are a question for a larger study. The address-authorship layer postdates the records of *the Casebook*. Re-extracting those records under it, and reporting what changes, is the subject of the next study.

## 8. Conclusion

A knowledge graph built over a contested field cannot dispense with a working taxonomy, and the Yue cases document what happens when that taxonomy is allowed to stand in for the classifications of the sources. Evidence remains open to inspection when the address used for retrieval, the category stated by a source and the decision admitting a passage to a comparison are held apart. Membership, rank and evidential basis then remain properties of each source's scheme, and a reclassification can be traced to a change in the evidence instead of appearing as a mere change of label. The same separation extends to a numerical parameter, where it marks the limits of verification at the level of the quotation. A quotation may support a record that nonetheless misleads. A citation preserves the passage, but says nothing of its scope, of the grounds for comparing it with others, or of how many records depend on one analysis.

What *the Casebook* leaves open is, for the most part, a sequence of tasks rather than a set of puzzles. For the address of a record, comparability decisions are already graph objects; the same form has to be extended to units of count, the scope of a passage and the dependence among records, and the revision test has to be run on it. The value schema has to admit structured counts, so that ‘five categories and eight values’ can be stored as the source states it, and the model, which already writes the scope of a number in free text, can be asked to fill the fields of a passport for a curator to confirm. Extraction across languages has to keep enumerations in the language of the source, since the losses in C1 arose in rendering a Chinese list into English, not in reading it. Beyond these lies the question for which the platform was built: once observations from Chinese, English and other scholarly literatures are comparable in this controlled sense, the relations among diagnostic features can be studied as data, and a classification can be tested against them.

Each of these steps can be specified now and checked against the records *the Casebook* already publishes; the first is to re-extract *the Casebook* sources under the address layer and report, on the same records, what changes. What limits them is no longer the design but the scale at which extraction, verification and re-verification can be repeated across a multilingual library, and it is at that scale that the coupling of language models and knowledge graphs ceases to be a convenience of retrieval and becomes an instrument of scholarship. The project intends to take these steps in this order. *The Casebook* is released so that others can check them, and carry them further, from the records rather than from this summary.

## Data and artefact availability

PaRAGraph Casebook, version 0.3 (public preview): https://www.sciencerag.win/casebook. Viewer code: MIT licence. Casebook text and curatorial annotations: CC BY 4.0. Quotations from third-party sources are excluded from the CC BY licence and are reproduced for scholarly discussion; the list of quoted works is given in the Casebook’s NOTICE file. Case bundles: C1 c1-internal-0.2, C2 c2-internal-0.3, C3 c3-internal-0.2, C4 c4-internal-0.1, inspectable in the viewer and distributed, with their manifests and hashes, in the source repository (https://github.com/dreamrandomhub/paragraph-casebook): for each case the records, their quotations, the curatorial annotations and, for C1–C3, the recorded results of native operations; for C4 the frozen export of 16 September 2026 and the three-variety subset. The evidence route of every statement cited above is given in the corresponding tab.

## Use of language models

Language models form part of the system studied here. Claims are extracted with Anthropic's Claude models (Sonnet and Opus), chosen after internal benchmarks that compared extraction quality across twelve models from several vendors. Verification within the pipeline (Section 3) uses OpenAI GPT models of the 5.5 generation and later, and answers in the platform's research chat are generated by MiniMax-M3. No chat output is cited as evidence in this article. All readings of sources, curatorial annotations, passport entries and case analyses reported here, as well as the research design itself, are the author's. The author compiled, reviewed and verified the text and takes responsibility for every argument, statement, source reading and conclusion.

## Acknowledgements

This publication is based upon my work as Scientific Research domain lead within COST Action CA24121 – Knowledge Graphs in the Era of Large Language Models (KGELL), supported by COST (European Cooperation in Science and Technology, https://www.cost.eu; https://kgell-cost.eu). This also forms my contribution as an affiliated researcher at the American University of Cyprus.